\documentclass[12pt]{article}
\usepackage{amsmath,amssymb, amsfonts, amsthm,amscd}
\usepackage[T2A]{fontenc}
\usepackage[cp1251]{inputenc}
\usepackage[russian,ukrainian,english]{babel}
\usepackage{geometry}
\usepackage{cite}

\newtheorem{Theorem}{Theorem}
\newtheorem{Corollary}{Corollary}

\newtheorem{Lemma}{Lemma}
\newtheorem{Proposition}{Proposition}

\newenvironment{LemmaProof}{\textbf{Proof. }}{\par\noindent\textbf{The Lemma is proved.}}
\newenvironment{TheoremProof}{\textbf{Proof. }}{\par\noindent\textbf{The Theorem is proved.}}
\newenvironment{PropProof}{\textbf{Proof. }}{\par\noindent\textbf{The Proposition is proved.}}

\title{{\Large \textbf{Interval colorings of edges of a multigraph}}}

\author{\textbf{\normalsize A.S. Asratian, R.R. Kamalian}}

\date{\small{A translation from Russian of the work of A.S. Asratian and R.R. Kamalian ''Interval colorings of
edges of a multigraph'', Applied Mathematics 5, Yerevan State
University, 1987, pp. 25--34.}}

\begin{document}

\maketitle

\bigskip

Let $G=(V(G),E(G))$ be a multigraph. The degree of a vertex $x$ in
$G$ is denoted by $d(x)$, the greatest degree of a vertex -- by
$\Delta(G)$, the chromatic index of $G$ -- by $\chi'(G)$. Let
$R\subseteq V(G)$.

An interval (respectively, continuous) on $R$ $t$-coloring of a
multigraph $G$ is a proper coloring of edges of $G$ with the colors
$1,2,\ldots,t$, in which each color is used at least for one edge,
and the edges incident with each vertex $x\in R$ are colored by
$d(x)$ consecutive colors (respectively, by the colors
$1,2,\ldots,d(x)$).

In this paper the problems of existence and construction of interval
or continuous on $R$ colorings of $G$ are investigated. Problems of
such kind appear in construction of timetablings without "windows".
Some properties of interval or continuous on $V(G)$ colorings were
obtained in \cite{Asr_Diss1, Kam_Dipl2}. Necessary and sufficient
conditions of the existence of a continuous on $V(G)$
$\Delta(G)$-coloring in the case when $G$ is a tree are obtained in
\cite{Caro3}. All non-defined concepts can be found in
\cite{Harary4, Zikov5}.

Let $\mathfrak{N}_t$ be the set of multigraphs $G$, for which there
exists an interval on $V(G)$ $t$-coloring, and
$\mathfrak{N}=\bigcup_{t\geq1}\mathfrak{N}_t$. For every
$G\in\mathfrak{N}$, let us denote by $w(G)$ and $W(G)$,
respectively, the least and the greatest $t$, for which there exists
an interval on $V(G)$ $t$-coloring of $G$. Evidently,
$\Delta(G)\leq\chi'(G)\leq w(G)\leq W(G)$.

\begin{Proposition}\label{prop1}
If $G\in\mathfrak{N}$ then $\chi'(G)=\Delta(G)$.
\end{Proposition}

\begin{PropProof}
Let us consider an interval on $V(G)$ $w(G)$-coloring of the
multigraph $G$. If $w(G)=\Delta(G)$ then $\chi'(G)=\Delta(G)$.
Assume that $w(G)>\Delta(G)$. Let us define the sets
$T(1),\ldots,T(\Delta(G))$, where $T(j)=\{i/\;i\equiv
j(mod(\Delta(G))),1\leq i\leq w(G)\}$, $j=1,\ldots,\Delta(G)$. Let
$E_j$ be the subset of edges of $G$ which are colored by colors from
the set $T(j)$, $j=1,\ldots,\Delta(G)$. Clearly, $E_j$ is a
matching. For each $j\in\{1,\ldots,\Delta(G)\}$, let us color the
edges of $E_j$ by the color $j$. We shall obtain a proper coloring
of edges of $G$ with $\Delta(G)$ colors. Hence,
$\chi'(G)=\Delta(G)$.
\end{PropProof}

\begin{Proposition}\label{prop2}
Let $G$ be a regular multigraph.

a) $G\in\mathfrak{N}$ iff $\chi'(G)=\Delta(G)$.

b) If $G\in\mathfrak{N}$ and $\Delta(G)\leq t\leq W(G)$ then
$G\in\mathfrak{N}_t$.
\end{Proposition}

\begin{PropProof}
The proposition (a) follows from the proposition \ref{prop1}. The
proposition (b) holds, since if $t>w(G)$ then an interval on $V(G)$
$(t-1)$-coloring can be obtained from an interval on $V(G)$
$t$-coloring by recoloring with the color $t-\Delta(G)$ all edges
colored by $t$.
\end{PropProof}

It is proved in \cite{Holyer6} that for a regular graph $G$, the
problem of deciding whether $\chi'(G)=\Delta(G)$ or
$\chi'(G)\neq\Delta(G)$ is $NP$-complete by R. Karp \cite{Karp7}. It
follows from here and from the proposition \ref{prop2} that for a
regular graph $G$, the problem of determining whether
$G\in\mathfrak{N}$ or $G\not\in\mathfrak{N}$ is $NP$-complete by R.
Karp.

\begin{Lemma}\label{lm1}
Let $G$ be a connected multigraph with a proper edge coloring with
the colors $1,\ldots,t$, and the edges incident with each vertex
$x\in V(G)$ are colored by $d(x)$ consecutive colors. Then
$G\in\mathfrak{N}$.
\end{Lemma}

\begin{LemmaProof}
Let $\alpha(e)$ be the color of the edge $e\in E(G)$. Without loss
of generality, we assume that $\min_e\alpha(e)=1$,
$\max_e\alpha(e)=t$. For the proof of the lemma it is suffice to
show, that if $t\geq3$, then each color $r$, $1<r<t$, is used for at
least one edge. Since $G$ is connected, then there exists a simple
path $P=(x_0,e_1,x_1,\ldots,x_{k-1},e_k,x_k)$ in it, where
$e_i=(x_{i-1},x_i)$, $i=1,\ldots,k$ and $\alpha(e_1)=t$,
$\alpha(e_k)=1$. If $\alpha(e_i)\neq r$, $i=1,\ldots,k$, let us
consider in $P$ the vertex $x_{i_0}$ with the greatest index,
satisfying the inequality $\alpha(e_{i_0})>r$. Then
$\alpha(e_{1+i_0})<r$. It follows from the condition of the lemma
that there is an edge incident with the vertex $x_{i_0}$ colored by
the color $r$.
\end{LemmaProof}

\begin{Theorem}\label{tm1}
Let $G$ be a connected graph without triangles. If
$G\in\mathfrak{N}$ then $W(G)\leq|V(G)|-1$.
\end{Theorem}

\textbf{Proof} by contrary. Assume that there exist connected graphs
$H$ in $\mathfrak{N}$ without triangles with $W(H)\geq|V(H)|$. Let
us choose among them a graph $G$ with the least number of edges.
Clearly, $|E(G)|>1$. Consider an interval on $V(G)$ $W(G)$-coloring
of $G$. The color of an edge $e$ is denoted by $\alpha(e)$, the set
$\{v\in V(G)/\;(u,v)\in E(G)\}$ -- by $I(u)$. Let $\mathfrak{M}$ be
the set of all simple paths with the initial edge colored by the
color $W(G)$ and the final edge colored by the color $1$. For each
$P\in\mathfrak{M}$ with the sequence $e_1,\ldots,e_t$ of edges,
$t\geq2$, let us set in correspondence the sequence
$\alpha(P)=(\alpha(e_1),\ldots,\alpha(e_t))$ of colors. Let us show
that there is a path $P_0$ in $\mathfrak{M}$ for which $\alpha(P_0)$
is decreasing.

Let $\alpha(e')=W(G)$, $e'=(x_0,x_1)$, and $d(x_1)\geq d(x_0)$.
Since $|E(G)|>1$, then $d(x_1)\geq2$. Let us construct the sequence
$X$ of vertices as follows:

Step 1. $X:=\{x_0,x_1\}$.

Step 2. Let $x_i$ be the last vertex in the sequence $X$. If
$I(x_i)\setminus X=\emptyset$ or $\alpha(x_i,y)>\alpha(x_{i-1},x_i)$
for each $y\in I(x_i)\setminus X$ then the construction of $X$ is
completed. Otherwise let us choose from $I(x_i)\setminus X$ the
vertex $x_{i+1}$, for which $\alpha(x_i,x_{i+1})=\min\alpha(x_i,y)$,
where the minimum is taken on all $y\in I(x_i)\setminus X$. Let us
bring $x_{i+1}$ in $X$ and repeat the step 2.

Suppose that $X$ is constructed and $X=\{x_0,x_1,\ldots,x_k\}$.
Clearly, $X$ defines a simple path
$P_0=(x_0,e_1,x_1,\ldots,x_{k-1},e_k,x_k)$, where
$e_i=(x_{i-1},x_i)$, $i=1,\ldots,k$. Let us show that
$\alpha(e_k)=1$.

Suppose $1<\alpha(e_k)<W(G)$. Let us define a graph $H$ as follows:
$$
H=\left\{
\begin{array}{ll}
G-x_k, & \textrm{if $d(x_k)=1$}\\
G-e_k, & \textrm{if $d(x_k)\geq2$}.\\
\end{array}
\right.
$$

Let us show that $H$ is connected. Assume the contrary. Then
$H=G-e_k$. Let $H_1$, $H_2$ be the connected components of $H$,
$x_{k-1}\in V(H_1)$, $x_k\in V(H_2)$, and $G_1$, $G_2$ be the
subgraphs of $G$ induced, respectively, by the subsets
$V(H_1)\cup\{x_k\}$ and $V(H_2)\cup\{x_{k-1}\}$. The coloring of the
graph $G$ induces the coloring of $G_i$ satisfying the conditions of
the lemma \ref{lm1}. Therefore $G_i\in\mathfrak{N}$, and, since
$|E(G_i)|<|E(G)|$ then $W(G_i)\leq|V(G_i)|-1$, $i=1,2$. It is not
difficult to check that $W(G)\leq W(G_1)+W(G_2)-1$. From here we
obtain the inequality $W(G)\leq|V(G)|-1$, which contradicts the
choice of $G$. Therefore $H$ is connected.

It follows from the lemma \ref{lm1} that the coloring of edges of
$H$ induced by the coloring of $G$ is an interval on $V(H)$
$W(G)$-coloring. Then $W(H)\geq W(G)\geq|V(G)|\geq|V(H)|$. The
obtained inequality contradicts the choice of $G$, because
$|E(H)|<|E(G)|$. Consequently, $\alpha(e_k)=1$.

Hence, we have constructed the path $P_0\in\mathfrak{M}$ for which
the sequence $\alpha(P_0)$ is decreasing.

Let us denote by $\vartheta$ the set of all shortest paths $P$ from
$\mathfrak{M}$, for which the sequence $\alpha(P)$ is decreasing.
Let $k$ be the length of paths in $\vartheta$.

Let us define the sets $\vartheta_1,\ldots,\vartheta_k$:
$\vartheta_1=\vartheta$, and $\vartheta_i$ is the subset of paths
from $\vartheta_{i-1}$ with the greatest color of the $i$-th edge,
$i=2,\ldots,k$.

Let us choose from $\vartheta_k$ some path
$P_1=(x_0,e_1,x_1,\ldots,x_{k-1},e_k,x_k)$. Let $A(i)=\{y\in
I(x_i)/\;\alpha(e_{i+1})<\alpha(x_i,y)<\alpha(e_i)\}$. Clearly,
$|A(i)|=\alpha(e_i)-\alpha(e_{i+1})-1$, $i=1,\ldots,k-1$. Let us
show that $A(i)\cap\{x_0,x_1,\ldots,x_k\}=\emptyset$,
$i=1,\ldots,k-1$. Suppose that there exist such $i_0$, $j_0$, that
either $x_{i_0}\in A(j_0)$ or $x_{j_0}\in A(i_0)$. Let us define the
path $P$ as follows. If $i_0\neq 0$, $j_0\neq k$, then
$P=(x_0,e_1,x_1,\ldots,x_{i_0},(x_{i_0},x_{j_0}),x_{j_0},\ldots,x_k)$.
If $i_0=0$, then $P=(x_1,e_1,x_0,(x_0,x_{j_0}),x_{j_0},\ldots,x_k)$.
If $j_0=k$, then
$P=(x_0,e_1,x_1,\ldots,x_{i_0},(x_{i_0},x_k),x_k,e_k,x_{k-1})$. In
all three cases the sequence $\alpha(P)$ is decreasing, and the
length of $P$ is less than the length of $P_1$, which contradicts
the choice of $P_1$. Therefore
\begin{equation}
A(i)\cap\{x_0,x_1,\ldots,x_k\}=\emptyset, \quad i=1,\ldots,k-1
\label{eq:1}
\end{equation}

Let us show that $A(i)\cap A(j)=\emptyset$, $1\leq i< j\leq k-1$.
Suppose that there exist such $i_0$, $j_0$, for which $1\leq i_0<
j_0\leq k-1$ and $A(i_0)\cap A(j_0)\neq\emptyset$.

Since there is no triangle in $G$ then $j_0-i_0\geq2$. Let $v\in
A(i_0)\cap A(j_0)$. Let us consider the path
$P=(x_0,e_1,x_1,\ldots,x_{i_0},(x_{i_0},v),v,(v,x_{j_0}),x_{j_0},\ldots,x_{k-1},e_k,x_k)$.
Clearly, the sequence $\alpha(P)$ is decreasing. If $j_0-i_0\geq 3$
then the length of $P$ is less than the length of $P_1$, and if
$j_0-i_0=2$ then $\alpha(x_{i_0},v)>\alpha(e_{1+i_0})$. In both
cases it contradicts the choice of $P_1$. Therefore
\begin{equation}
A(i)\cap A(j)=\emptyset, \quad 1\leq i< j\leq k-1 \label{eq:2}
\end{equation}

From (\ref{eq:1}) and (\ref{eq:2}), it follows that
$$
|V(G)|\geq\bigg|\bigcup_{i=1}^{k-1}A(i)\bigg|+k+1=k+1+\sum_{i=1}^{k-1}|A(i)|=
k+1+\sum_{i=1}^{k-1}(\alpha(e_i)-\alpha(e_{i+1})-1)=
$$

$=k+1+W(G)-1-(k-1)=1+W(G)$.

It contradicts the choice of $G$.

\textbf{The Theorem is proved.}

\begin{Corollary}\label{cor1}
If $G$ is a connected bipartite graph, and $G\in\mathfrak{N}$, then
$W(G)\leq|V(G)|-1$.
\end{Corollary}

Let $G=(V_1(G),V_2(G),E(G))$ be a bipartite multigraph. Let us
denote by $w_1(G)$ and $W_1(G)$, respectively, the least and the
greatest $t$, for which there exists an interval on $V_1(G)$
$t$-coloring of $G$. Evidently, $W_1(G)=|E(G)|$.

\begin{Theorem}\label{tm2}
For any $t$, $w_1(G)\leq t\leq W_1(G)$, there exists an interval on
$V_1(G)$ $t$-coloring of the multigraph $G$.
\end{Theorem}

\textbf{Proof} by induction on $|V_1(G)|$.

If $|V_1(G)|=1$ then the proposition of the theorem is true. Suppose
that the proposition of the theorem is true for all $G'$ with
$|V_1(G')|=p$. Suppose that $|V_1(G)|=p+1$, and assume there exists
an interval on $V_1(G)$ $t$-coloring of $G$, $w_1(G)\leq t< W_1(G)$.
Among vertices of $V_1(G)$ which are incident with edges colored by
the color $t$, let us choose a vertex $x_1$ with the smallest
degree. There is an edge $e_1$ colored by the color $t+1-d(x_1)$
which is incident with the vertex $x_1$.

1) If there exists an edge different from $e_1$ and colored by the
color $t+1-d(x_1)$, then, by recoloring $e_1$ with the color $t+1$
we shall obtain an interval on $V_1(G)$ $(t+1)$-coloring of $G$.

2) Let $e_1$ be the unique edge colored by the color $t+1-d(x_1)$,
and $s$ be the maximum color which is used for more than one edge.
Clearly, $1\leq s<t<|E(G)|$.

2a) Let $t+1-d(x_1)<s<t$. Let us recolor each edge with the color
$i$, where $i=t+1-d(x_1),\ldots,s$, by the color $i+t-s$, and let us
recolor each edge with the color $i$, where $i=s+1,\ldots,t$, by the
color $(i+t-s)(modt)+t-d(x_1)$. In the obtained interval on $V_1(G)$
$t$-coloring, among that vertices from $V_1(G)$ which are incident
with edges colored by $t$ (there are more than 1 such edges), we
shall choose a vertex $x_2$ with the smallest degree and recolor the
incident with it edge with the color $t+1-d(x_2)$ by the color
$t+1$. We shall obtain an interval on $V_1(G)$ $(t+1)$-coloring of
$G$.

2b) Let $1\leq s<t+1-d(x_1)$. Removing $x_1$ from $G$, we shall
obtain a multigraph $G'$ with an interval on $V_1(G')$
$(t-d(x_1))$-coloring. Clearly, $|E(G')|=|E(G)|-d(x_1)$ and
$t-d(x_1)<|E(G')|=W_1(G')$. By the assumption of induction there
exists an interval on $V_1(G')$ $(t+1-d(x_1))$-coloring of $G'$. We
shall color the edges incident with the vertex $x_1$ by the colors
$t+2-d(x_1),\ldots,t+1$ and obtain an interval on $V_1(G)$
$(t+1)$-coloring of $G$.

\textbf{The Theorem is proved.}

\bigskip

In the work \cite{Even8} in terms of timetables the
$NP$-completeness was proved for the problem of a $3$-coloring of a
bipartite graph with preassignments in one part. A bipartite graph
$H=(V_1(H),V_2(H),E(H))$ with $\Delta(H)=3$ is given, where the set
$V_1(H)$ contains no pendent vertex, and, for each $x\in V_1(H)$, a
set $T(x)$ is preassigned, $T(x)\subseteq\{1,2,3\}$,
$|T(x)|=d_H(x)$. The required is to determine does there exist a
proper coloring of edges of $H$ with the colors $1,2,3$, at which
the edges incident with each vertex $x\in V_1(H)$ are colored by
colors from the set $T(x)$.

\begin{Theorem}\label{tm3}
For a bipartite multigraph with the greatest degree $3$ of a vertex,
the problem of deciding whether a $3$-coloring, continuous on one
part, exists or not, is $NP$-complete.
\end{Theorem}

\begin{TheoremProof}
Let $H'$ be a graph isomorphic to the graph $H$, $V(H)\cap
V(H')=\emptyset$, and to each vertex $y\in V(H)$ a vertex $y'\in
V(H')$ corresponds. Let us construct a bipartite multigraph $G_1$ as
follows.

For each $y\in V_2(H)$, connect the vertices $y$ and $y'$ with one
edge if $d_H(y)=2$, and with two parallel edges if $d_H(y)=1$.
Clearly, $\Delta(G_1)=3$.

Set
$$
\begin{array}{l}
T(y'):=T(y), T(y):=T(y) \textrm{ for each } y\in V_1(H),\\
T(y'):=\{1,2,3\}, T(y):=\{1,2,3\} \textrm{ for each } y\in V_2(H).
\end{array}
$$

Let $V_{ij}=\{y\in V(G_1)/T(y)=\{i,j\}\}$, $1\leq i<j\leq 3$.

Let us define a bipartite multigraph $G$ as follows:
$$
\begin{array}{l}
V(G)=V(G_1)\cup\{x_1/x\in V_{23}\}\cup\{x_1,x_2/x\in V_{13}\},\\
E(G)=E(G_1)\cup\{(x,x_1)/x\in V_{23}\}\cup\{(x,x_1),(x_1,x_2)/x\in
V_{13}\}.
\end{array}
$$

Clearly, a $3$-coloring of $H$ with preassignments in $V_1(H)$
exists if and only if a $3$-coloring of edges of $G_1$ exists at
which the edges incident with each vertex $x\in V(G_1)$ are colored
by colors from the set $T(x)$. Such coloring of edges of $G_1$
exists if and only if a $3$-coloring of $G$ continuous on $V(G)$
exists. It is not difficult to check that the collection of degrees
of vertices of $V_1(G)$ coincides with the collection of degrees of
vertices of $V_2(G)$. Therefore a continuous on $V(G)$ $3$-coloring
of $G$ exists if and only if a continuous on $V_1(G)$ $3$-coloring
of $G$ exists.
\end{TheoremProof}

\begin{Theorem}\label{tm4}
Let $G=(V_1(G),V_2(G),E(G))$ be a bipartite multigraph. If for each
edge $(x,y)$, where $x\in V_1(G)$, the condition $d(x)\geq d(y)$
holds, then $G$ has a continuous on $V_1(G)$ $\Delta(G)$-coloring.
\end{Theorem}

\begin{TheoremProof}
Let $V_1(G)=\{x_1,\ldots,x_p\}$, $d(x_1)\geq\ldots\geq d(x_p)$, and
already a proper coloring of edges incident with the vertices
$x_1,\ldots,x_n$ $(n\geq 1)$ is constructed so that the edges
incident with the vertex $x_i$ are colored by the colors
$1,\ldots,d(x_i)$, $i=1,\ldots,n$. If $n<p$ and with the vertex
$x_{n+1}$ the edges $(x_{n+1},y(1)),\ldots,(x_{n+1},y(d(x_{n+1})))$
are incident, then, sequentially for each $j=1,\ldots,d(x_{n+1})$ do
as follows. If the color $j$ is absent in the vertex $y(j)$, then we
shall color the edge $(x_{n+1},y(j))$ by the color $j$. Otherwise a
color $k$ is absent in $y(j)$, $1\leq k\leq d(x_{n+1})$. We shall
recolor the longest path consisting of edges colored by $j$ and $k$
with the initial vertex $y(j)$ and we shall color the edge
$(x_{n+1},y(j))$ by the color $j$.
\end{TheoremProof}

\begin{Corollary}\label{cor2}
If
$$
\min_{x\in V_1(G)}d(x)\geq \max_{y\in V_2(G)}d(y),
$$
then $G$ has a continuous on $V_1(G)$ $\Delta(G)$-coloring.
\end{Corollary}

\begin{Proposition}\label{prop3}
The problem of deciding whether a proper coloring of edges of a
bipartite multigraph with the fixed number of edges of each color
exists is $NP$-complete.
\end{Proposition}

\begin{PropProof}
Let $G=(V_1(G),V_2(G),E(G))$ be a bipartite multigraph with
$\Delta(G)=3$. Set $n_i=|\{x/\;x\in V_1(G), d(x)\geq i\}|$,
$i=1,2,3$. Clearly, a continuous on $V_1(G)$ $3$-coloring of $G$
exists if and only if there exists a proper coloring of edges of $G$
with the colors $1,2,3$, at which by each color $i$ $n_i$ edges are
colored, $i=1,2,3$. Therefore, the proposition \ref{prop3} follows
from the theorem \ref{tm3}.
\end{PropProof}

Some sufficient conditions for the existence of a proper coloring of
edges of a bipartite multigraph with the fixed number of edges
colored by each color are found in \cite{Folkman9, deWerra10, Asr11,
Asr12}.
\bigskip\bigskip

\begin{flushright}
\textit{The Chair of Mathematical Cybernetics of YSU, \\
The Computing Centre of the Academy of \\
Sciences of Armenian SSR and YSU}
\end{flushright}

\bigskip\bigskip

\begin{center}
\textbf{Resume (in Armenian)}
\end{center}

\bigskip

Let $G=(V_1(G),V_2(G),E(G))$ be a bipartite multigraph, and
$R\subseteq V_1(G)\cup V_2(G)$. A proper coloring of edges of $G$
with the colors $1,\ldots,t$ is called interval (respectively,
continuous) on $R$, if each color is used for at least one edge and
the edges incident with each vertex $x\in R$ are colored by $d(x)$
consecutive colors (respectively, by the colors $1,\ldots,d(x))$,
where $d(x)$ is a degree of the vertex $x$. We denote by $w_1(G)$
and $W_1(G)$, respectively, the least and the greatest values of
$t$, for which there exists an interval on $V_1(G)$ coloring of the
multigraph $G$ with the colors $1,\ldots,t$.

In the paper the following basic results are obtained.

\textbf{Theorem 2.} For an arbitrary $k$, $w_1(G)\leq k\leq W_1(G)$,
there is an interval on $V_1(G)$ coloring of the multigraph $G$ with
the colors $1,\ldots,k$.

\textbf{Theorem 3.} The problem of recognition of the existence of a
continuous on $V_1(G)$ coloring of the multigraph $G$ is
$NP$-complete.

\textbf{Theorem 4.} If for any edge $(x,y)\in E(G)$, where $x\in
V_1(G)$, the inequality $d(x)\geq d(y)$ holds then there is a
continuous on $V_1(G)$ coloring of the multigraph $G$.

\textbf{Theorem 1.} If $G$ has no multiple edges and triangles, and
there is an interval on $V(G)$ coloring of the graph $G$ with the
colors $1,\ldots,k$, then $k\leq|V(G)|-1$.

\end{document}